\documentclass[fleqn,twoside]{article}
\usepackage{espcrc2}

\usepackage{graphicx}
\usepackage[figuresright]{rotating}

\newcommand{\AmS}{{\protect\the\textfont2
  A\kern-.1667em\lower.5ex\hbox{M}\kern-.125emS}}

\title{Deeply Virtual Neutrino Scattering at Leading Twist\thanks{Presented at Neutrino
Oscillations Workshop (NOW 2006), Otranto, Lecce, Italy, September 9-16 2006}}

\author{C. Corian\`o \address[MCSD]{Dipartimento di Fisica, Universit\`a di Lecce,
        and INFN sez. Lecce, \\
        via per Arnesano, 73100 Lecce, Italy.}%
        M. Guzzi\address{Dipartimento di Fisica, Universit\`a di Lecce\\
        via per Arnesano, 73100 Lecce, Italy.}
		  }

\begin{document}

\begin{abstract}
We illustrate the generalization of results concerning
exclusive electromagnetic processes in the deeply virtual limit (DVCS) to the case of
the weak interactions.We briefly describe the derivation 
of the differential cross section for neutrino-nucleon
reactions mediated by the neutral current and the amplitude for similar reactions mediated by the charged currents.
\vspace{1pc}
\end{abstract}

\maketitle

\section{Introduction}

Exclusive processes mediated by the weak force are
an area of investigation which will gather, we believe, a wide interest
in the forthcoming years due to the various experimental proposals to detect
neutrino oscillations at intermediate energy using neutrino factories and superbeams,
\cite{Schaefer1}.
Also, neutrino scattering on nucleons in the deeply virtual kinematics
is important in order to draw a better picture of the hadronic interactions mediated by the weak currents.
This is one of the simplest processes, among all, described by the Generalized Parton
distribution functions (GPDFs) which unify the parton model description of both inclusive ed exclusive
processes \cite{Ji,Radyushkin}. Exclusive processes in the region of small momentum transfer
(few GeVs) are well described by hadronic wave functions - also called "distribution amplitudes" -
using the ERBL formalism of Efremov-Radyushkin-Brodsky-Lepage (see \cite{CorianoLi} \cite{Sterman}
for an overview),
while inclusive processes at high energy are described by ordinary parton distributions.
GPDs are the appropriate tool to describe the so called ``Generalized Bjorken region''
which is the region where, beside Bjorken's x variable, additional scaling variables appear.
Nonforward parton distributions are the simplest constructs belonging to this class
and describe true light-cone correlators in the nucleon state, not built by unitarity as
ordinary parton distributions, since the optical theorem does not hold. An asymmetry scaling parameter,
in this case, $\xi$, defined below, characterizes the longitudinal momentum exchange
between the initial and final momenta of the nucleon. The study of the electromagnetic interaction
in this kinematical region, the Deeply Virtual Compton Scattering (DVCS) region, by now amounts to 
several hundreds of papers written in the last few years. Our objective, in this contribution, is to 
briefly outline the extension to the electroweak sector of these findings.

\section{DVNS kinematics}
\begin{figure}[t]
{\par\centering \resizebox*{8cm}{!}{\includegraphics{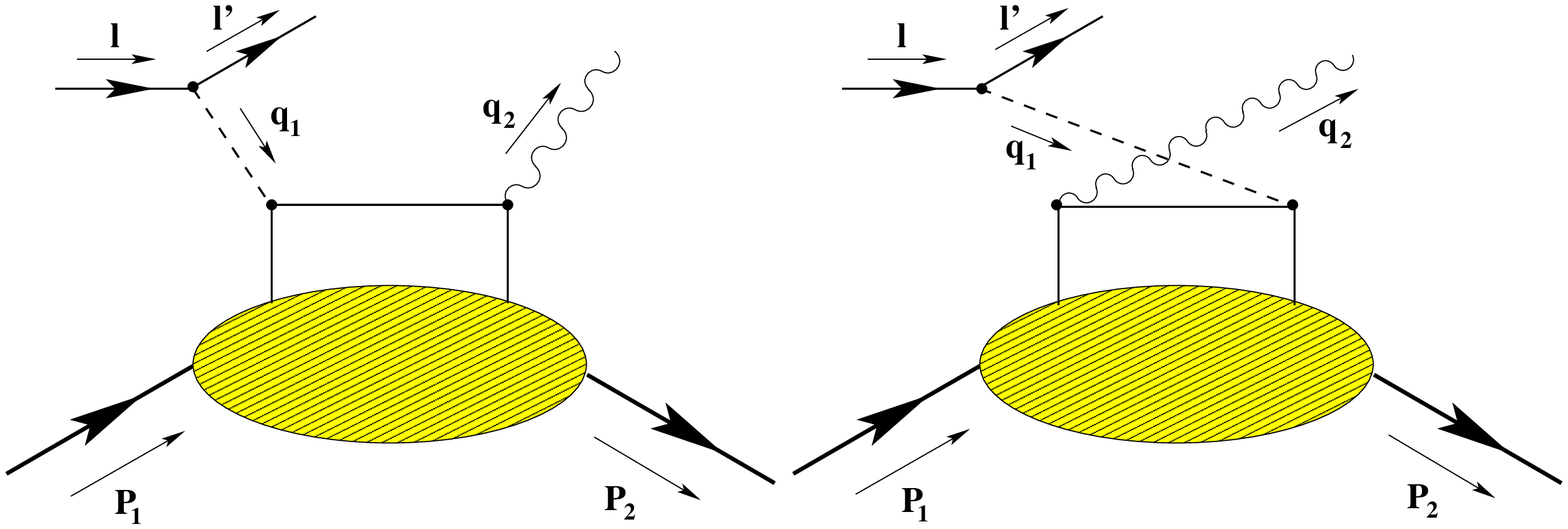}} \par}
\caption{Leading hand-bag diagrams for the process}
\label{DVCS_1.eps}
\end{figure}

A pictorial description of the process we are going to illustrate
is given in Fig~\ref{DVCS_1.eps} where a neutrino of momentum $l$
scatters off a nucleon of momentum $P_1$ by an interaction with a neutral current;
from the final state a photon and a nucleon emerge, of momenta $q_2$ and $P_2$ respectively,
while the momenta of the final lepton is $l'$
\begin{equation}
P_{1,2}= \bar{P} \pm \frac{\Delta}{2}\,\,\,\,\,\,\,\, q_{1,2}= \bar{q} \mp \frac{\Delta}{2},
\end{equation}
with $-\Delta=P_2-P_1$ being the momentum transfer. Clearly
\begin{equation}
\bar{P}\cdot \Delta=0,\,\,\,\,\, t=\Delta^2 \,\,\,\,\,\,\, \bar{P}^2=M^2 - \frac{t}{4}
\end{equation}
and $M$ is the nucleon mass. There are two scaling variables which are identified in the process,
since 3 scalar products  can grow large in the generalized Bjorken limit:
$\bar{q}^2$, $\Delta\cdot q$, $\bar{P}\cdot \bar{q}$.

The momentum transfer $t=\Delta^2$ is a small parameter in the process.
Momentum asymmetries between the initial and the final state nucleon are measured
by two scaling parameters, $\xi$ and $\eta$, related to ratios of the former invariants
\begin{equation}
\xi=-{\bar{q}^2\over 2 \bar{P}\cdot {\bar q}} \,\,\,\,\,\,\,\,\,\,\,\,
\eta={\Delta\cdot \bar{q}\over 2 \bar{P}\cdot \bar{q}},
\end{equation}
where $\xi$ is a variable of Bjorken type, expressed in terms of average
momenta rather than nucleon and Z-boson momenta.
The standard Bjorken variable $x= - q_1^2/( 2 P_1\cdot q_1)$ is trivially
related to $\xi$ in the $t=0$ limit.
In the DIS limit $(P_1=P_2)$ $\eta=0$ and $x=\xi$, while in the DVCS limit
$\eta=\xi$ and $x=2\xi/(1 +\xi)$, as one can easily deduce from the relations
\begin{equation}
q_1^2=\left(1 +\frac{\eta}{\xi}\right) \bar{q}^2 +\frac{t}{4},\,\,\,
q_2^2=\left(1 -\frac{\eta}{\xi}\right) \bar{q}^2 +\frac{t}{4}.
\end{equation}
We introduce also the inelasticity parameter $y=P_1\cdot l/(P_1\cdot q_1)$
which measures the fraction of the total energy that is transferred to the final state photon.

\section{The Amplitudes}

The Compton amplitude is described by the following correlator
involving neutral and charged currents
\begin{small}
\begin{eqnarray}
T_{\mu \nu}=i\int d^{4}x e^{i qx} \langle P_2|T\left(J_{\nu}^{\gamma}(x/2)
J_{\mu}^{W^{\pm},Z_{0}}(-x/2)\right)|P_1\rangle.
\end{eqnarray}
\end{small}
The parameterizations of the non-forward light cone correlators in terms of
GPD's is of the form given by Ji at leading twist \cite{Ji}
\begin{small}
\begin{eqnarray}
&&\int\frac{d\lambda }{(2\pi)}e^{i\lambda z}\langle P'|\overline {\psi}
\left(-\frac{\lambda n}{2}\right)\gamma^\mu \psi\left(\frac{\lambda n}{2}\right)|P\rangle=
\nonumber\\\nonumber\\
&&H(z,\xi,\Delta^2)\overline{U}(P')\gamma^\mu U(P)\nonumber\\
&&\hspace{1cm}+ E(z,\xi,\Delta^2)\overline{U}(P')\frac{i\sigma^{\mu\nu}
\Delta_{\nu}}{2M} U(P) + .....\nonumber\\
&&\int\frac{d\lambda }{(2\pi)}e^{i\lambda z}\langle P'|\overline {\psi}
\left(-\frac{\lambda n}{2}\right)\gamma^\mu \gamma^5 \psi\left(\frac{\lambda n}{2}\right)|P\rangle=
\nonumber\\\nonumber\\
&&\tilde{H}(z,\xi,\Delta^2)\overline{U}(P')\gamma^\mu\gamma^5 U(P)\nonumber\\
&&\hspace{1cm}+ \tilde{E}(z,\xi,\Delta^2)\overline{U}(P')\frac{\gamma^5 \Delta^{\mu}}{2M}U(P) + .....
\end{eqnarray}
\end{small}
which have been expanded in terms of functions $H,E, \tilde{H},
\tilde{E}$ which are the GPDs and the ellipses are meant to denote
the higher-twist contributions. It is interesting to observe that the amplitude
is still described by the same light-cone correlators as in the electromagnetic
case (vector, axial vector) but now parity is not conserved.

\begin{figure}[t]
{\par\centering \resizebox*{6cm}{!}{\rotatebox{-90}{\includegraphics{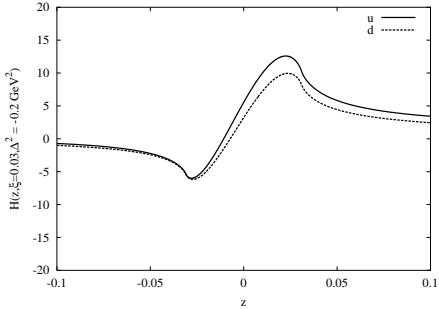}}} \par}
\caption{GPD's $H_u$ and $H_d$ generated by the diagonal parton distribution
at an initial $0.26$ GeV$^2$}
\label{Hud}
\end{figure}

In our calculations we use a simplified model for the GPD's
where the $\Delta^2$ dependence can be factorized as follows \cite{Schaefer2}
for a generic GPD
\begin{equation}
H^{i}(z,\xi,\Delta^2,Q^2)=F^{i}_{1}\,(\Delta^2)q^{i}(z,\xi,Q^2),
\end{equation}
where $F^{i}_{1}\,(\Delta^2)$ is a form factor, and the construction of the input distributions,
in correspondence of an input scale $Q_0$, is performed using
\begin{eqnarray}
&&\hspace{-0.5cm}q(z,\xi,Q^2)=\int_{-1}^{1}dx'\int_{-1+|x'|}^{1-|x'|}dy'\delta(x'+\xi y'-z)\nonumber\\
&&\hspace{0.5cm}\times f(y',x',Q^2).
\label{reduction}
\end{eqnarray}
As an example we show a plot of some GPDs in Fig~(\ref{Hud}), for the up and the down quarks as a function
of the variable $z$ (a generalization of Bjorken-x, with $-1 < z < 1$) and at a fixed value of the 
variables $\Delta$ and $\xi$.

\section{The cross sections}

One possible differential cross section to study, in analogy to the DVCS case, is the following
\begin{eqnarray}
&&\frac{d\sigma}{dx d Q^2 d|\Delta^2| d\phi_{r}}=
\frac{y}{Q^2}\frac{d\sigma}{dx dy d|\Delta^2| d\phi_{r}}=\nonumber\\
&&\frac{x y^2}{8\pi Q^4}
\left(1+ \frac{4M^2 x^2}{Q^2}\right)^{-\frac{1}{2}}|{\cal M}_{fi}|^2,
\end{eqnarray}
where $\phi_{r}$ is the angle between the lepton and the hadron
scattering planes. $|{\cal M}_{fi}|^2$, as a functions of the invariants of the
process, can be found in \cite{DVNS1}.
\begin{figure}
{\centering \resizebox*{7cm}{!}{\rotatebox{-90}{\includegraphics{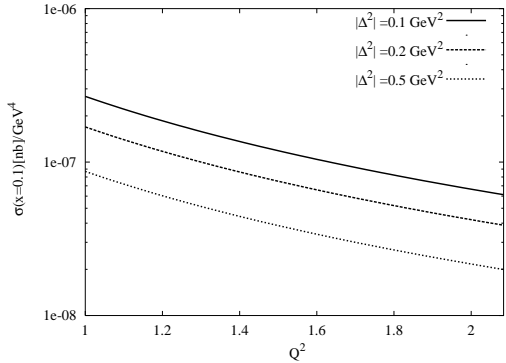}}} \par}
\caption{DVCS cross section at $x=0.1$ and center of mass energy $M E=10$ GeV$^2$.}
\label{set2.ps}
\end{figure}

\section{Amplitudes in the charged sector}
We have also analized the expressions for the correlators of
the charged electroweak current and derived their expressions in terms of the GDPs \cite{DVNS2}.
These are formally given by the correlators
\begin{small}
\begin{eqnarray}
&&\hspace{-1cm}T^{W^{+}}_{\mu \nu}=i\int d^4 x \frac{e^{iqx}x^{\alpha}U_{u d}}{2\pi^2(x^2-i\varepsilon)^2}\langle P_2|
\left[i S_{\mu\alpha\nu\beta}\left(\tilde{O}_{u d}^{\beta}+O_{u d}^{5 \beta}\right)
\right.\nonumber\\
&&\hspace{2cm}\left.+\varepsilon_{\mu\alpha\nu\beta}\left(O_{u d}^{\beta}+\tilde{O}_{u d}^{5 \beta}\right)\right] |P_1\rangle,
\nonumber\\
&&\hspace{-1cm}T^{W^{-}}_{\mu \nu}=i\int d^4 x \frac{e^{iqx}x^{\alpha}U_{d u}}{2\pi^2(x^2-i\varepsilon)^2}\langle P_2|
\left[-i S_{\mu\alpha\nu\beta}\left(\tilde{O}_{d u}^{\beta}+O_{d u}^{5 \beta}\right)
\right.\nonumber\\
&&\hspace{2cm}\left.-\varepsilon_{\mu\alpha\nu\beta}\left(O_{d u}^{\beta}+\tilde{O}_{d u}^{5 \beta}\right)\right] |P_1\rangle\,,
\nonumber\\
\end{eqnarray}
\end{small}
where the partonic content is extracted from the $O_{d u}$ operators while $S_{\mu\alpha\nu\beta}$ 
denote suitable combinations of Dirac gamma matrices. We refer to \cite{DVNS2} for more details.
We have suppressed the $x$-dependence of the operators in the former equations.
The identification of the relevant partonic description in terms of GPDs has been obtained by us \cite{DVNS2}
using the leading-twist decomposition worked out in \cite{blumlein}. Related work, using a different
approach, can be found also in \cite{Psaker}.

\section{Conclusions}

We have presented an extension of the standard DVCS process to the case
of one neutral current exchange, describing the scattering of a neutrino
off a proton in the parton model.
We have described the leading twist behaviour of the cross section;
we have found that this is comparable to other typical
neutrino cross sections.
The process is the natural generalization of DIS with neutral currents and relies on
the notion of Generalized Parton Distributions, new constructs in the parton model
which have received considerable attention in recent years.
From the theoretical and experimental viewpoints
the study of these processes is of interest,
since very little is known of the neutrino interaction at intermediate energy
in these more complex kinematical domains.

\end{document}